# Nanostructured BiVO$_4$ Photoanodes Fabricated by Vanadium-infused Interaction for Efficient Solar Water Splitting

Amar K. Salih[1,2], Abdul Zeeshan Khan[3], Qasem A. Drmosh[4], Tarek A. Kandiel[3,4*], Mohammad Qamar[4], Tahir Naveed Jahangir[3], Cuong Ton-That[1, *], Zain H. Yamani[2,4]

[1]School of Mathematical and Physical Sciences, University of Technology Sydney, Ultimo, New South Wales 2007, Australia
[2]Physics Department, King Fahd University of Petroleum and Minerals, Dhahran, 31261, Saudi Arabia
[3]Chemistry Department, King Fahd University of Petroleum and Minerals, Dhahran, 31261, Saudi Arabia
[4]Center of Hydrogen and Energy Storage (IRC-HES), King Fahd University of Petroleum and Minerals, Dhahran 31261, Saudi Arabia

*Corresponding authors.
 E-mail address:  Cuong.Ton-That@uts.edu.au ; tarek.kandiel@kfupm.edu.sa

**ABSTRACT**

Bismuth vanadate (BiVO$_4$) has emerged as a highly prospective material for photoanodes in photoelectrochemical (PEC) water oxidation. However, current limitations with this material lie in the difficulties in producing stable and continuous BiVO$_4$ layers with efficient carrier transfer kinetics, thereby impeding its widespread application in water splitting processes. This study introduces a new fabrication approach that yields continuous, monoclinic nanostructured BiVO$_4$ films, paving the way for their use as photoanodes in efficient PEC water oxidation for hydrogen production under solar light conditions. The fabrication involves the intercalation of vanadium (V) ions into Bi$_2$O$_3$ films at 450°C. Upon interaction with V ions, the film undergoes a transformation from tetragonal Bi$_2$O$_3$ to monoclinic scheelite BiVO$_4$. This synthesis method enables the fabrication of single monoclinic phase BiVO$_4$ films with thicknesses up to 270 nm. The engineered monoclinic BiVO$_4$ film, devoid of any pinholes that could cause carrier loss, exhibits a robust photocurrent of 1.0 mA/cm$^2$ at 1.23 V$_{RHE}$ in a neutral electrolyte, without requiring additional modifications or doping. Moreover, we demonstrate that the incorporation of a cobalt phosphate (Co–Pi) co-catalyst into the BiVO$_4$ photoanode significantly enhances the lifetime of photogenerated holes by a factor of nine, resulting in a further elevation of the photocurrent to 2.9 mA/cm$^2$. This remarkable PEC enhancement can be attributed to the surface state passivation by the Co–Pi co-catalyst. Our fabrication approach opens up a new facile route for producing large-scale, highly efficient BiVO$_4$ photoanodes for PEC water splitting technology.

## 1. Introduction

Solar energy conversion to electrochemical energy offers a viable avenue for producing hydrogen through the photoelectrochemical (PEC) water splitting process.[1] This PEC process heavily relies on suitable semiconductor photoanodes that can facilitate efficient light absorption and fast charge transport at the semiconductor/electrolyte interface (SEI). Metal oxide semiconductors such as ZnO, $WO_3$, $Fe_2O_3$, $TiO_2$, and $BiVO_4$ are promising candidates for practical PEC water oxidation owing to their facile preparation, low fabrication cost and photo-oxidative capability.[2, 3] Among these oxides, monoclinic $BiVO_4$ stands out due to its narrow bandgap of 2.4 eV and a favorable valence band edge position for oxygen evolution. These characteristics enable efficient water oxidation with a recently reported solar-to-hydrogen (STH) efficiency of 9.2%.[4-6]

$BiVO_4$ photoanodes have been theoretically predicted to achieve 7.5 mA/cm² at 1.23 $V_{RHE}$ (voltage with respect to reversible hydrogen electrode) under AM1.5 G illumination (100 mW/cm²);[7] however, experimentally reported photocurrent densities fall well below this anticipated value. This disparity primarily arises from the low utilization of photogenerated electron-hole pairs in the PEC processes. Extensive research efforts have been devoted to improving water oxidation efficiency through the optimization of various $BiVO_4$ fabrication techniques. Presently, $BiVO_4$ films are commonly produced using physical vapor deposition, which offers advantages such as uniformity, large photoanodes, and adjustable film thickness. However, this method typically produces structure-less films with limited surface areas available for the photo-oxidative reaction, leading to low water oxidation efficiencies. Photocurrents in the range of 0.01–0.5 mA/cm² at 1.23 $V_{RHE}$ are typically achieved even after fine-tuning the deposition parameters of $BiVO_4$ photoanodes or employing co-sputtering of $Bi_2O_3$ and V targets.[8-10] Conversely, chemical fabrication methods can yield nanostructured films with a high specific surface area for $BiVO_4$ photoanodes. However, a major drawback of these methods is the formation of a discontinuous BiVO$_4$ layer,[11-14] which can lead to significant PEC efficiency loss due to solution-mediated reduction of photogenerated carriers.[15] Figure 1 illustrates the primary pathways of carrier loss on BiVO$_4$ photoanodes deposited on a typical fluorine-doped tin oxide (FTO) substrate. The water oxidation process is inherently slow, which allows photogenerated holes to recombine with electrons before participating in the oxidation reaction. This recombination commonly occurs at defects within the bulk of the BiVO$_4$ film (process a) or on its surface (process b). Additionally, back reduction of the oxidation products and electron back injection through solution-mediated recombination at voids or exposed regions of the film (process c) have been identified as substantial carrier loss mechanisms.[15, 16] These pathways collectively limit the efficiency and stability of BiVO$_4$ photoanodes in water splitting applications. Addressing these challenges is imperative for enhancing the performance of BiVO$_4$-based water splitting systems.

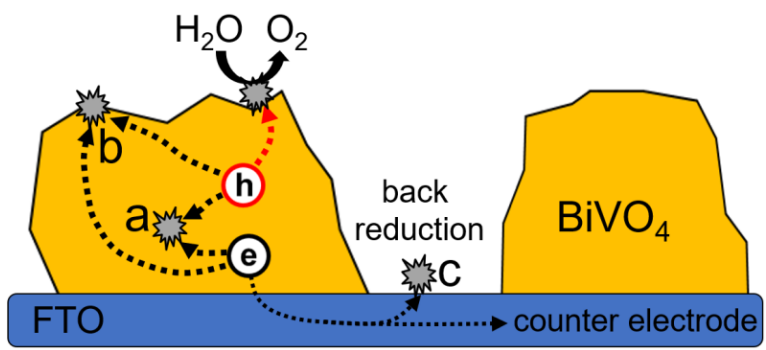

**Figure 1.** Schematic illustrating the primary reaction pathways responsible for the decline in PEC efficiency in water oxidation. Three main routes are: electron-hole recombination at bulk defects (process a) or surface defects (b), and solution-mediated back reduction occurring through exposed regions of the FTO substrate (c). Arrows show the trajectories of electrons and holes through the BiVO$_4$ film and FTO substrate.

This work presents a new method for fabricating continuous, nanostructured BiVO$_4$ films without pinholes, establishing them as highly efficient photoanodes for PEC water oxidation in hydrogen production. The BiVO$_4$ photoanode in the monoclinic scheelite phase is fabricated by intercalating vanadium (V) ions into a sputtered Bi$_2$O$_3$ film at 450 °C. The analysis evaluates the quality of the intercalated films and their PEC performance over film thicknesses from 60 to 1100 nm. The optimized BiVO$_4$ photoanode in this fabrication approach is shown to exhibit an exceptional and stable photocurrent density of up to 1.0 mA/cm$^2$ under neutral pH and in the absence of a hole scavenger. The outstanding performance of the BiVO$_4$ photoanode arises from the elimination of solution-mediated recombination through exposed FTO regions (process c) and the optimization of the charge transfer path through the photoanode thickness. Moreover, we demonstrate that the incorporation of cobalt phosphate (Co–Pi) catalyst can effectively suppress surface recombination (process b) and extend the lifetime of photogenerated holes by nine times, resulting in an outstanding photocurrent up to 2.9 mA/cm$^2$ at 1.23 V$_{RHE}$.

## 2. Experimental Details

*2.1 Preparation of BiVO$_4$ films*

BiVO$_4$ and Co–Pi/BiVO$_4$ films were fabricated through a process involving V intercalation and Co–Pi electrodeposition, as illustrated in Scheme 1. Bi$_2$O$_3$ films, with thickness ranging from 60 to 1100 nm, were deposited on the FTO substrate (8 Ω/sq resistivity, from Sigma-Aldrich), with a geometrical area of 1.0 cm$^2$, using metallic Bi target (99.995% purity) and a direct current (DC) reactive sputtering magnetron (Nanomaster, NSC-4000). Throughout the deposition process, the chamber's base pressure remained at $9 \times 10^{-6}$ torr, and the working pressure was maintained at $3 \times 10^{-3}$ torr within an atmosphere of oxygen and argon gases (1:1 ratio of oxygen to argon). The thickness of the Bi$_2$O$_3$ films was controlled by adjusting the deposition time. Following this, 30 µL of a solution containing 0.15 M vanadyl acetylacetonate ($C_{10}H_{14}O_5V$, $VO(acac)_2$) dissolved in $C_2H_6OS$ was drop-casted onto the $Bi_2O_3$ film surface. The films were then annealed for 2 hours at 450 °C (ramping rate of 2 °C/min) in an ambient atmosphere. At this temperature, the decomposition of $VO_{(acac)2}$ led to the production of $VO_x$ species,[17] facilitating their diffusion into the $Bi_2O_3$ film. The intercalated film was subsequently agitated in a 1.0 M NaOH solution for 10 min to remove any access of $VO_x$ on the surface. To further enhance the PEC performance, the $BiVO_4$ films were modified with a Co–Pi co-catalyst using photo-assisted electrochemical deposition. This deposition was conducted at 1.6 $V_{RHE}$ for 3 minutes using 0.5 mM $Co(NO_3)_2$ in 0.1 M $H_2KO_4P$ (phosphate-buffered saline, PBS) solution with the pH maintained at 7.

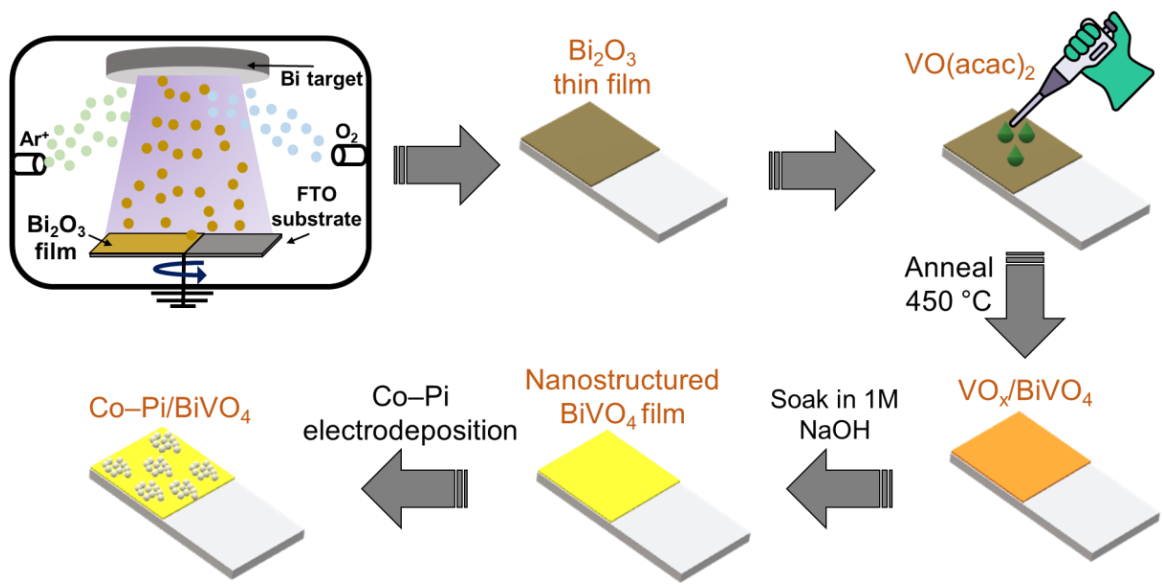

**Scheme 1.** Schematic illustration of the fabrication steps for $BiVO_4$ and Co–Pi/$BiVO_4$ photoanodes.

*2.2 Structural and PEC characterization*

X-ray diffraction (XRD) analysis was carried out using a Rigaku Miniflex 600 X-ray Diffractometer. An Agilent Cary 5000 UV-Vis-NIR spectrometer, operating in transmission mode, was employed to assess the optical properties of the films. Film surface morphology and cross-sections were examined using Tescan Lyra-3 field emission scanning electron microscopy (FE-SEM). Chemical analysis via X-ray photoelectron spectroscopy (XPS) was performed using the ESCALAB250Xi Thermo Fisher spectrometer.

Photoelectrochemical (PEC) performance was tested in a three-electrode configuration using an Autolab potentiostat (PGSTAT 302N). The intercalated $BiVO_4$ films were used as the working electrode, while a Pt coil and Ag/AgCl (3.0 KCl) were used as the counter and reference electrodes, respectively. A neutral solution of 0.5 M PBS was used as an electrolyte in a homemade cell with a quartz window. To investigate the impact of hole scavengers on PEC performance, 0.5 M PBS + 0.5 M $Na_2SO_3$ aqueous solution was used as an electrolyte to minimize surface electron-hole pair recombination on the $BiVO_4$ photoanode.[18] A solar simulator 1002 SunLitewas used to simulate solar conditions with a light intensity of 100 mW/cm$^2$. The reversible hydrogen electrode potential ($E_{RHE}$) was calculated from the applied potential ($E_{Ag/AgCl}$) using the Nernst equation:[19]

$$E_{RHE} = E_{Ag/AgCl} + 0.0591 \cdot pH + E^o_{Ag/AgCl} \qquad \text{Eq. (1)}$$

where $E^o_{Ag/AgCl}$ is the standard electrode potential, which is 0.198 V. The chronoamperometry response for the photoanode was recorded to calculate the incident photon to current efficiency (IPCE) at 1.23 $V_{RHE}$. Collimated monochromatic LED light sources (Thorlabs) with wavelengths between 375 to 565 nm were used, and light intensity was calibrated using an FDS100-CAL photodiode. For the photoelectrochemical impedance spectroscopy (PEIS), the same Autolab potentiostat setup was employed. Nyquist plots were measured at a biased potential of 0.6 $V_{RHE}$ over the frequency range from 100 kHz to 0.1 Hz. Finally, Mott-Schottky analysis was performed at 10 kHz in a potential window from 0.3 – 1.4 $V_{RHE}$ under dark conditions, and only the linear part of the $1/C^2$ versus potential plot was considered.

## 3. Results and discussion

### 3.1 Structural, morphological, optical, and compositional analyses

Fig. 2(a) presents the XRD patterns of both V-intercalated $BiVO_4$ and $Bi_2O_3$ films for comparison. Detailed XRD analysis, including the indexation of all diffraction peaks in the $BiVO_4$ and $Bi_2O_3$ films with thicknesses ranging from 60 to 1100 nm, is provided in Fig. S1(a). The $Bi_2O_3$ film exhibits dominant XRD peaks corresponding to (220), (013), and (123) reflections of the tetragonal crystal structure. Upon V intercalation into the 60-nm and 270-nm thick films, these $Bi_2O_3$ peaks completely disappear, indicating a complete phase transformation into monoclinic scheelite $BiVO_4$. The V-intercalated $BiVO_4$ films all display clearly identifiable peaks (011), (121), (040), (211) and (161) of the monoclinic phase (JCPDS no. 14–0688); however, residual $Bi_2O_3$ tetragonal phase is observed in 520, 800, and 1100 nm thick films, likely due to incomplete diffusion of V ions into these thick $Bi_2O_3$ films. No distinct Co–Pi XRD peaks are detectable in the $BiVO_4$ film after the deposition of the Co–Pi catalyst [Figure S1(b)]. XPS analysis of the V-intercalated $BiVO_4$ films, presented in Fig. S2, confirms the formation of $BiVO_4$ after the V intercalation, with the presence of anticipated photoemission signatures corresponding to Bi, V, and O. The strong V 2p XPS signal after $Ar^+$ ions etching for 120 seconds provides further support that V ions are incorporated throughout the thickness of the film, extending down to the substrate. The Bi 4f spectrum reveals a doublet-peak consisting of Bi $4f_{7/2}$ and $4f_{7/2}$ components at the binding energies of 158.6 eV and 164.0 eV, consistent with reported values for with $BiVO_4$.[20] Furthermore, the successful incorporation of the Co–Pi catalyst on the film surface is confirmed by the Co 2p and P 2p XPS spectra in Fig. S2(e, f).

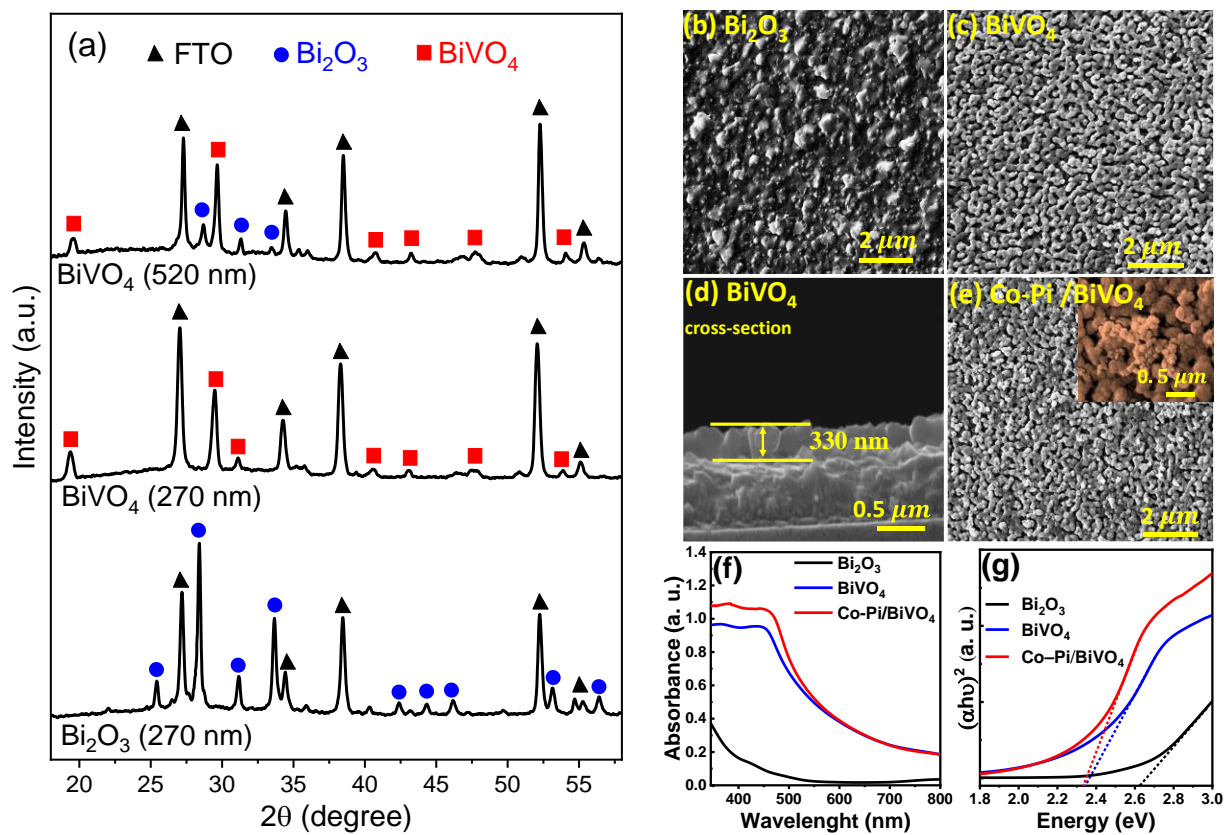

**Figure 2.** (a) XRD patterns of BiVO$_4$ and Bi$_2$O$_3$ films with thicknesses of 270 nm and 520 nm. Black triangles represent the FTO substrate, while blue circles and red squares indicate reflections from tetragonal Bi$_2$O$_3$ and monoclinic BiVO$_4$ phases, respectively. Complete phase transition to monoclinic BiVO$_4$ occurs in 270-nm thick film, while a residual Bi$_2$O$_3$ phase is present in the 520-nm thick film. (b-e) SEM images of the 270-nm Bi$_2$O$_3$ film, its corresponding BiVO$_4$, and Co–Pi/BiVO$_4$ films. Both BiVO$_4$ and Co–Pi/BiVO$_4$ films exhibit continuous, nanostructured morphology. (f) Optical absorbance spectra of Bi$_2$O$_3$, BiVO$_4$, and Co–Pi/BiVO$_4$ films showing an ~ 80% increase in optical absorption at wavelengths below 500 nm after the transformation to BiVO$_4$, while the Co–Pi deposition induces marginal absorption rise. (g) Tauc plots yielding $E_g$ = 2.64 ± 0.05 eV for Bi$_2$O$_3$ and $E_g$ = 2.36 ± 0.03 eV for BiVO$_4$ and Co–Pi/BiVO$_4$ films.

The surface morphology of a typical 270 nm $Bi_2O_3$ film and corresponding $BiVO_4$ film after the V intercalation are displayed in the top-view and cross-section SEM images in Fig. 2(b-e). The initial $Bi_2O_3$ film exhibits a rough, island-like surface structure. However, following the V intercalation, the film is transformed into a continuous, nanostructured $BiVO_4$ film with no discernible pinholes. The $BiVO_4$ film has a granular surface texture, characterized by densely packed grains with diameters of about 100 nm. This pinhole-free and nanostructured nature, coupled with the achievement of a pure monoclinic phase in the $BiVO_4$ film, plays a crucial role in enhancing the PEC oxidation efficiency, as discussed above. Notably, the thickness of the $BiVO_4$ film increases slightly from 270 to 330 nm following the V incorporation, as shown in Fig. 2(d). The SEM image of the Co–Pi/$BiVO_4$ film in Fig. 2(e) reveals the presence of Co–Pi catalyst nanoparticles with diameters of ~ 30 - 50 nm on the film surface. Notably, the film morphology remains unaffected by the Co–Pi electrodeposition process. The energy dispersive X-ray (EDX) spectrum of the Co–Pi/$BiVO_4$ film and the SEM image of Co–Pi nanoparticles, presented in Fig. S3, confirm the successful deposition of the Co–Pi particles onto the $BiVO_4$ film. Upon the V intercalation, the film exhibits a substantial increase of approximately 80% in optical absorption at wavelengths below 500 nm compared to the $Bi_2O_3$ film, as shown in Figure 2(f), consistent with the formation of monoclinic $BiVO_4$. The addition of Co–Pi co-catalyst does not result in any noticeable impact on the optical absorption of the film. The optical bandgap ($E_g$) of the films, determined using the Tauc plot in Fig. 2(g), is $E_g$ = 2.64 eV ± 0.05 eV for the $Bi_2O_3$ film and $E_g$ = 2.36 ± 0.03 eV for the $BiVO_4$ and Co–Pi/$BiVO_4$ films; the latter value is in agreement with the bandgap of monoclinic $BiVO_4$.[21] The optical transmission spectrum of the $BiVO_4$ film, presented in Figure S4, shows a transmittance of < 10% for wavelengths below 450 nm and a broad edge that allows this film to absorb well into the visible range.

*3.2 Photoelectrochemical water splitting performance*

The linear sweep voltammetry (LSV) characteristics of the intercalated BiVO$_4$ in Fig. 3(a) show an increase in current density under illumination, with zero current in the dark. The observed photocurrent peak at around 0.3 V$_{RHE}$ is associated with the redox transition between V$^{4+}$ and V$^{5+}$ states within BiVO$_4$ photoanodes.[22] The BiVO$_4$ film ($t$ = 60 nm) exhibits a relatively low photocurrent (0.15 mA/cm$^2$ at 1.23 V$_{RHE}$), likely due to its low thickness and reduced crystallinity; these factors contribute to diminished light absorption and increased charge carrier scattering.[23] The BiVO$_4$ photoanode ($t$ = 270 nm) exhibits the highest PEC performance, with a photocurrent density of 1.0 mA/cm$^2$ at 1.23 V$_{RHE}$, measured under neutral pH and in the absence of a hole scavenger. This level of photocurrent ranks among the highest reported for uncatalyzed BiVO$_4$ photoanodes recorded under comparable conditions, as summarized in Table S1. The excellent PEC performance is attributed to the high film quality achieved through our new fabrication approach. The BiVO$_4$ photoanodes with greater Bi$_2$O$_3$ thicknesses ($t$ = 520, 800, and 1100 nm) exhibit lower photocurrent densities, as shown in the inset of Fig. 3(a). In these thicker films, the possible formation of a BiVO$_4$/Bi$_2$O$_3$ heterojunction could cause holes in the BiVO$_4$ layer to migrate towards the underlying Bi$_2$O$_3$ layer due to the band alignment and the built-in electric field at the heterojunction interface, consequently reducing the availability of holes for water oxidation.[24] Another factor contributing to the lower photocurrent densities in the thick films could be an increased film resistance, thus constraining the availability of photogenerated holes at the BiVO$_4$ surface.[25, 26] Fig. 3(b) presents an LSV comparison between the BiVO$_4$ and Co–Pi/BiVO$_4$ photoanodes in PBS electrolyte, alongside the BiVO$_4$ photoanode in PBS + Na$_2$SO$_3$ electrolyte. The comparison reveals an increase in the optimum BiVO$_4$ photocurrent to around 3.1 mA/cm$^2$ at 1.23 V$_{RHE}$ after the addition of the hole scavenger Na$_2$SO$_3$. This suggests a heightened frequency of $SO_3^{2-}$ oxidation by photogenerated holes on the BiVO$_4$ surface, compared with the water oxidation process.[27] Furthermore, in comparison to other BiVO$_4$ electrodes synthesized via chemical fabrication methods,[28, 29] the photocurrent attained during the hole-scavenger test by this BiVO$_4$ photoanode is notably higher (3.1 mA/cm$^2$). This observation suggests an effective suppression of charge recombination that may occur at the exposed areas of the substrate (see Figure 1). Following the Co–Pi incorporation into the optimized BiVO$_4$ photoanode, the photocurrent reaches 2.9 mA/cm$^2$ at 1.23 V$_{RHE}$, with an onset potential of 0.25 V$_{RHE}$; this performance closely approaches the level achieved with the hole scavenger [Fig. 3(b)]. The enhanced photocurrent can be attributed to holes in the valence band of BiVO$_4$ transferring to the Co–Pi co-catalyst, where they participate in the oxidation of Co$^{2+}$ ions to Co$^{3+}$ and potentially further to Co$^{4+}$, which are the highly active species driving the water oxidation.[30] The onset potential of the photoanodes, determined from the linear extrapolation of LSV curves, is 0.91 and 0.29 V for BiVO$_4$ and Co–Pi/BiVO$_4$, respectively. This large cathodic shift indicates diminished surface charge recombination facilitated by the Co–Pi co-catalyst. Further analysis of the LSV curves obtained for the BiVO$_4$ and Co–Pi/BiVO$_4$ photoanodes under chopped light illumination (Fig. S5) reveal diminished anodic spikes with increasing applied potential. This observation suggests that the applied positive potential effectively diminishes electron concentration from surface states, thereby minimizing charge recombination and simultaneously enhancing the kinetics of hole transfer at the SEI during the water oxidation process.

The stability of the BiVO$_4$ and Co–Pi/BiVO$_4$ photoanodes was examined over a 2-hour duration, as depicted in Fig. 3(c). The results reveal that the virtually unchanged photocurrent at 1.0 mA/cm$^2$ during this period. However, for the Co–Pi/BiVO$_4$ photoanode, the photocurrent gradually decreases from 2.9 to 2.5 mA/cm$^2$ after 2 hours. This decline is attributed to the generation of high photocurrent, causing partial blockage of the electrode surface due to continuous bubble formation. Chopped light chronoamperometry was also conducted for the $BiVO_4$ and Co–Pi/$BiVO_4$ photoanodes at 1.23 $V_{RHE}$ under standard illumination conditions, as shown in Fig. 3(d). The consistent photocurrents observed during the cycles of light-on and light-off for these photoanodes confirm their fast photo-response and highly stable photocurrent. Furthermore, the anodic spike after sudden illumination for the Co–Pi/$BiVO_4$ photoanode is considerably higher than for $BiVO_4$, indicating a greater accumulation of holes on the Co–Pi/$BiVO_4$ surface compared to the bare $BiVO_4$ surface.

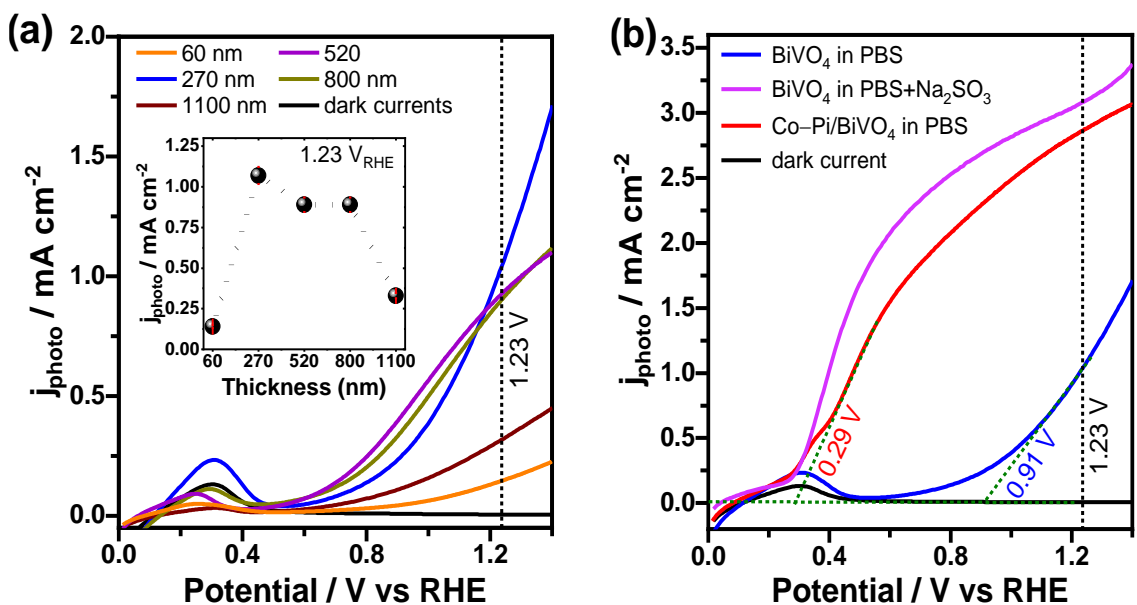

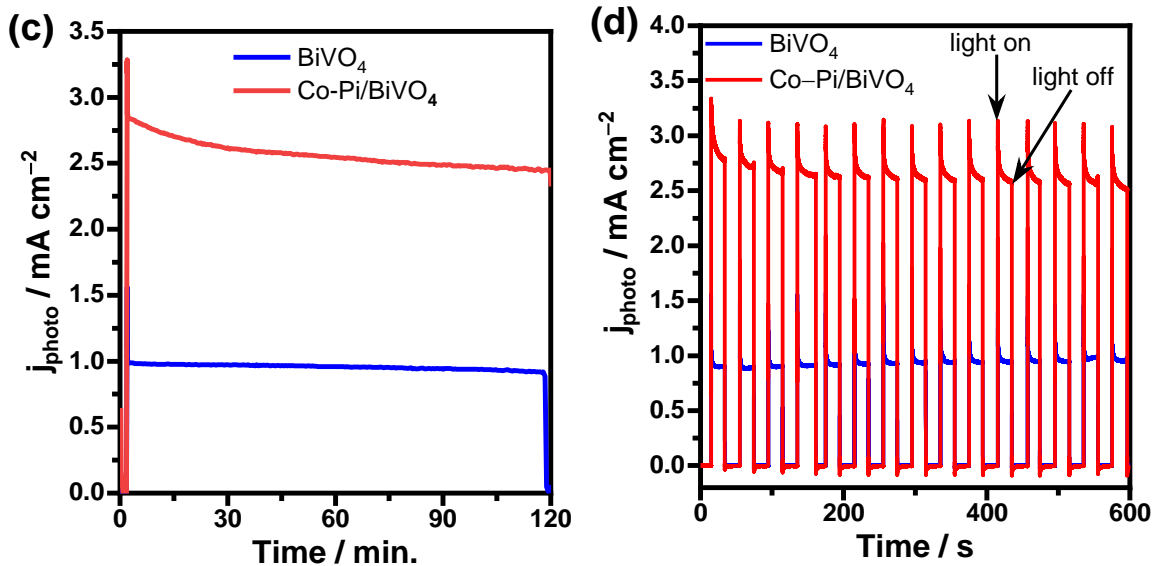

**Figure 3.** (a) LSV curves of the $BiVO_4$ photoanodes derived from $Bi_2O_3$ films with varying thicknesses between 60 and 1100 nm. The inset shows the highest photocurrent of 1.0 mA/cm² at 1.23 $V_{RHE}$ for the 270 nm thick photoanode. (b) LSV curves of the $BiVO_4$ and Co–Pi/$BiVO_4$ photoanodes in PBS electrolyte, and $BiVO_4$ photoanode in PBS + $Na_2SO_3$ electrolyte. After Co–Pi loading, the photocurrent increases from around 1.0 to 2.9 mA cm$^{-2}$ at 1.23 $V_{RHE}$. (c) Endurance testing performed on $BiVO_4$ and Co–Pi/$BiVO_4$ photoanodes, demonstrating significant stability over 2 hours of continuous illumination. (d) Chopped light stability testing for the photoanodes, revealing consistent and fast responses during light modulation, with no electro-current detected in the absence of light.

The performance of the intercalated $BiVO_4$ photoanode was further examined via incident photon-to-current efficiency (IPCE) across the wavelength range from 375 to 565 nm, as shown in Fig. 4(a). The IPCE of the $BiVO_4$ photoanode reaches approximately 9% at 405 nm, which is higher than reported values of 3% deposited $BiVO_4$ films.[28, 31] The significant increase in the IPCE value is attributed to the uniformity of the $BiVO_4$ film and complete coverage of the FTO substrate, reducing the carrier loss due to back reduction as illustrated in Fig. 1. As expected, the IPCE increases to ~ 25% after the addition of the Co–Pi co-catalyst, which is consistent with the LSV results in Fig 3(b). The transient photocurrent (TPC) shows that both $BiVO_4$ and Co–Pi/$BiVO_4$ photoanodes exhibit anodic and cathodic photocurrent spikes, as shown in Fig. S6. The anodic spike ($I_m$) is observed following sudden illumination, resulting from the accumulation of holes on surface states and ultimately reaching a stable photocurrent ($I_s$). The transient decay time ($\tau$) of the Co–Pi/BiVO₄ and BiVO₄ photoanodes is evaluated to gain further insights into the lifetime of the trapped holes. A parameter $D$ is calculated from the TPC response using:[32]

$$D = \frac{I_t - I_s}{I_m - I_s} \qquad \text{Eq. (2)}$$

where $I_t$ is the time-dependent photocurrent. The decay time, determined from the plot of ln(D) versus t shown in Fig. 4(b), is found to be $\tau = 0.10 \pm 0.01$ and $0.90 \pm 0.05$ s for the BiVO₄ and Co–Pi/BiVO₄ photoanodes, respectively. This nine-fold increase in the decay time for the Co–Pi/BiVO₄ photoanode indicates photogenerated holes remain available much longer (9 times longer) for water oxidation. Such a prolonged lifetime of photogenerated holes can be attributed to the passivation of BiVO₄ surface states by Co–Pi.

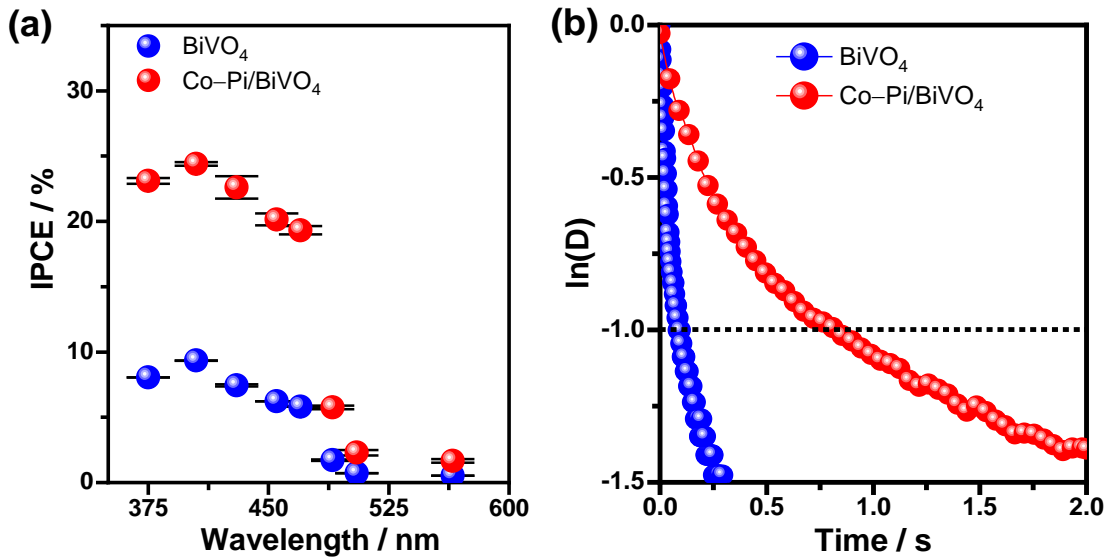

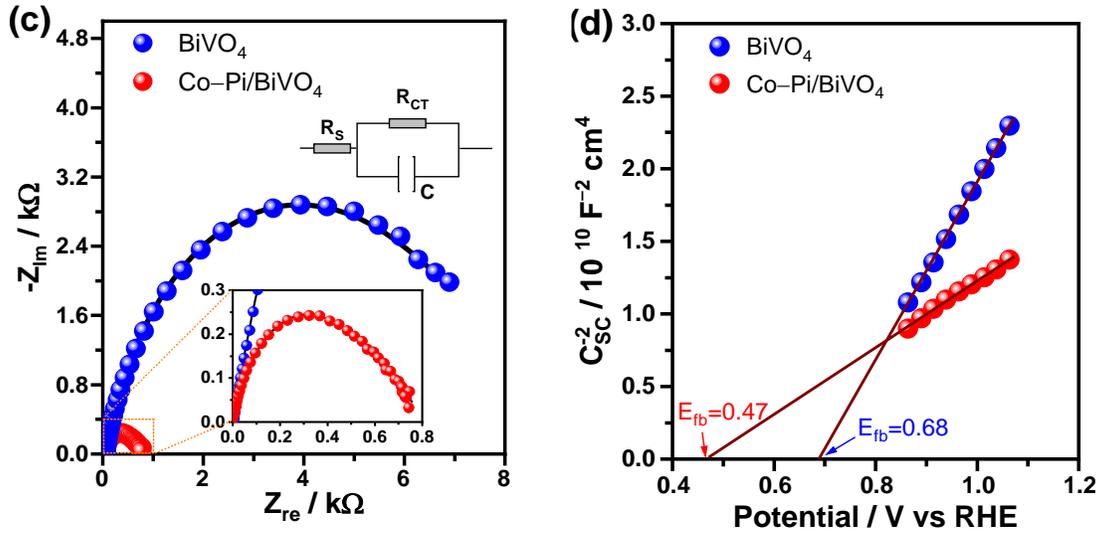

**Figure 4.** (a) IPCE for BiVO$_4$ and Co–Pi/BiVO$_4$ photoanodes measured at various wavelengths between 375 and 565 nm. The black lines indicate estimated uncertainties. (b) Plot of ln(D) versus time for the photoanodes, yielding $\tau$ = 0.10 ± 0.01 and 0.90 ± 0.05 s for BiVO$_4$ and Co–Pi/BiVO$_4$, respectively, indicating a remarkable enhancement in hole lifetime due to surface passivation by Co–Pi. (c) Nyquist plots and the equivalent circuit model (inset) for the BiVO$_4$ and Co–Pi/BiVO$_4$ photoanodes at 0.6 V$_{RHE}$. (d) Mott-Schottky plots for the photoanodes under dark conditions. The positive slope indicates *n*-type characteristics, with a cathodic shift of $E_{fb}$ by -0.21 V$_{RHE}$.

The PEIS method was used to investigate the charge transfer kinetics in the electrodes, as shown in Fig. 4(c). The Nyquist plots for the BiVO$_4$ and Co–Pi/BiVO$_4$ photoanodes are fitted to the equivalent circuit (depicted in the inset) for comparative analysis of the charge transfer resistance at the SEI. After the incorporation of Co–Pi onto the BiVO$_4$ photoanode, the charge transfer resistance ($R_{ct}$) decreases dramatically from 7.88 kΩ for the BiVO$_4$ to 0.73 kΩ for the Co–Pi/BiVO$_4$ photoanode. This substantial reduction indicates a notable improvement in mitigating the surface state recombination by Co–Pi. The markedly lower $R_{ct}$ for the Co–Pi/BiVO$_4$ photoanode further corroborates that that Co–Pi can effectively passivate surface states, thereby extending the lifetime of photogenerated holes engaged in the water oxidation process. Figure 4(d) shows the Mott-Schottky plots for the BiVO$_4$ and Co–Pi/BiVO$_4$ photoanodes. The presence of a distinct positive slope in the linear part of the plots indicates the manifestation of *n*-type characteristics of the photoanode films. The donor density, calculated from the Mott-Schottky slope, is $N_d = 5.4 \times 10^{20}$ cm$^3$ and $14.3 \times 10^{20}$ cm$^3$ for BiVO$_4$ and Co–Pi/BiVO$_4$, respectively. These values demonstrate an order of magnitude higher than similar BiVO$_4$ photoanodes prepared by direct deposition.[33-35] After Co–Pi passivation on the BiVO$_4$ electrode, a noticeable cathodic shift in the flat-band potential ($E_{fb}$) is observed, with its value decreasing from 0.68 V$_{RHE}$ to 0.47 V$_{RHE}$. This large cathodic shift of -0.21 V$_{RHE}$ is highly advantageous as it promotes the flow of electrons across the circuit toward the counter electrode, and as a result, this leads to a reduction in the onset potential for anodic photocurrent.[36]

**Conclusion**

Our study introduces a new approach for producing continuous, nanostructured BiVO$_4$ films for enhanced PEC water oxidation in hydrogen production. This approach involves intercalating V ions into a sputtered Bi$_2$O$_3$ film through a gradual annealing process. The optimized BiVO$_4$ photoanode, with a thickness of 270 nm, exhibits outstanding PEC performance with a strong, robust photocurrent density of 1.0 mA/cm$^2$ at 1.23 V$_{RHE}$ in a neutral electrolyte, notably without the necessity for a hole scavenger. This performance ranks among the best recorded for uncatalyzed BiVO$_4$ photoanodes measured under comparable conditions. The absence of pinholes in our intercalated BiVO$_4$ film effectively suppresses solution-mediated recombination at FTO regions exposed to the electrolyte, thus increasing the availability of photogenerated holes for water oxidation. Furthermore, the addition of the Co–Pi co-catalyst to the BiVO$_4$ photoanode facilitates the passivation of surface defects, enhancing the hole lifetime by a remarkable 9-fold. This enhancement results in an exceptional photocurrent of 2.9 mA/cm$^2$ at 1.23 V$_{RHE}$ under standard illumination conditions.

**Acknowledgments**

The authors would like to acknowledge the support provided by the Center of Hydrogen and energy storage (HES) and the Deanship of Scientific Research (DSR) at King Fahd University of Petroleum & Minerals (KFUPM) for funding this work through project No# DF181021. This work was supported under Australian Research Council (ARC) Discovery Project funding scheme (project DP210101146).

**Supplementary data**

Supplementary data to this paper can be found online at xxx.

Supporting Information

# Nanostructured BiVO$_4$ Photoanodes Fabricated by Vanadium-infused Interaction for Efficient Solar Water Splitting

Amar K. Salih[1,2], Abdul Zeeshan Khan[3], Qasem A. Drmosh[4], Tarek A. Kandiel[3,4,*], Mohammad Qamar[4], Tahir Naveed Jahangir[3], Cuong Ton-That[1,*], Zain H. Yamani[2,4]

[1]School of Mathematical and Physical Sciences, University of Technology Sydney, Ultimo, New South Wales 2007, Australia
[2]Physics Department, King Fahd University of Petroleum and Minerals, Dhahran, 31261, Saudi Arabia
[3]Chemistry Department, King Fahd University of Petroleum and Minerals, Dhahran, 31261, Saudi Arabia
[4]Center of Hydrogen and Energy Storage (IRC-HES), King Fahd University of Petroleum and Minerals, Dhahran 31261, Saudi Arabia

*Corresponding authors.
E-mail address: Cuong.Ton-That@uts.edu.au ; tarek.kandiel@kfupm.edu.sa

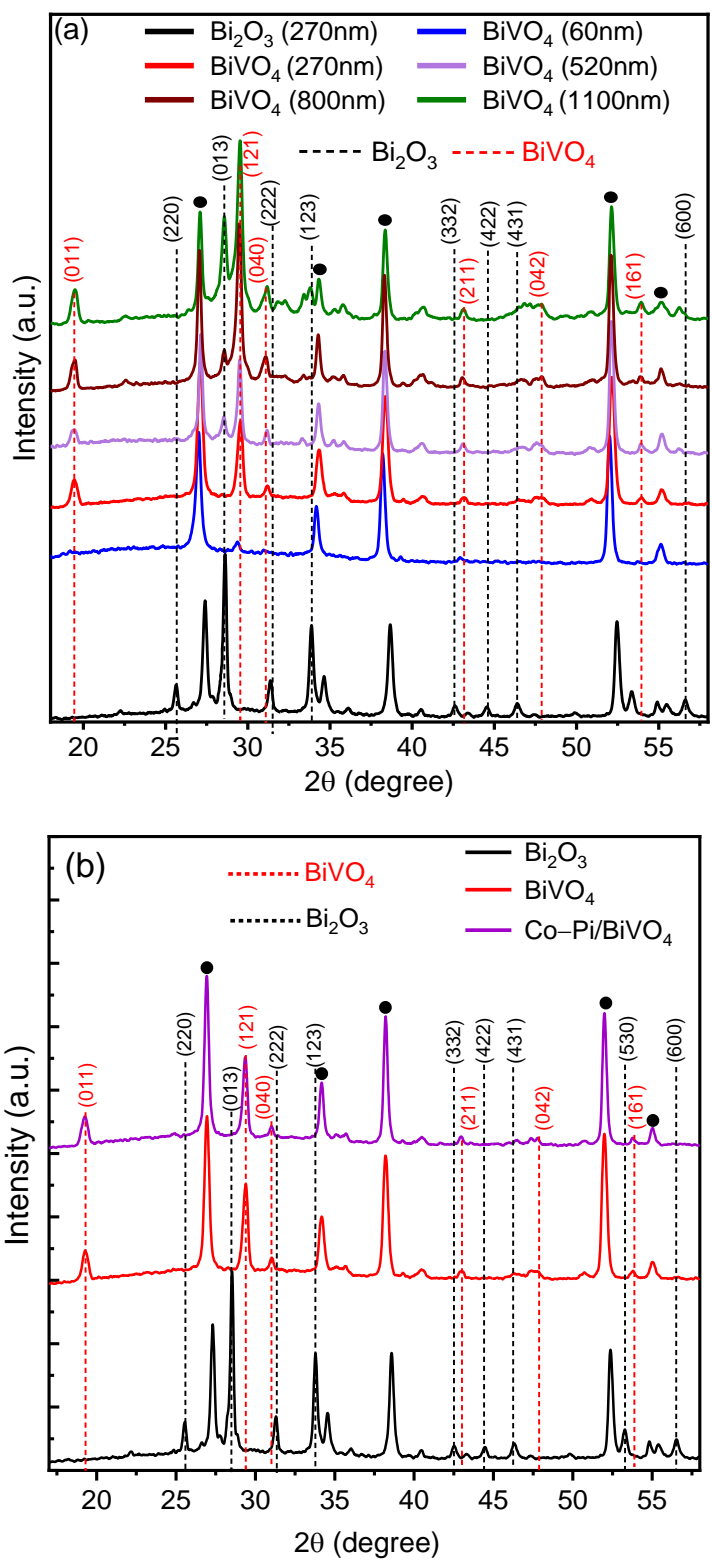

**Figure S1.** (a) XRD patterns of BiVO$_4$ films obtained from V intercalation into Bi$_2$O$_3$ films with thicknesses ranging from 60 to 1100 nm, and the XRD pattern of the pristine Bi$_2$O$_3$ film prior to the V intercalation. Black dots correspond to the FTO substrate, while black and red dashed lines indicate reflections from tetragonal Bi$_2$O$_3$ and monoclinic BiVO$_4$ phases, respectively. The Bi$_2$O$_3$ film exhibits dominant XRD peaks corresponding to (220), (013), and

(123) reflections of the tetragonal structure. These $Bi_2O_3$ peaks completely disappear after V intercalation in the thinner $BiVO_4$ films (60 and 270 nm thick), revealing a complete phase transformation into a monoclinic $BiVO_4$ phase. The transformed $BiVO_4$ films display clearly identifiable peaks (011), (121), (040), (211) and (161) of the monoclinic phase (JCPDS no. 14–0688). However, residual $Bi_2O_3$ tetragonal phase is detected in the films with thicknesses of 520, 800, and 1100 nm thick films. (b) XRD patterns of the $BiVO_4$ and Co–Pi/$BiVO_4$ films fabricated from a 270 nm thick $Bi_2O_3$ film, alongside the XRD pattern of tetragonal $Bi_2O_3$ for comparison. No distinct Co–Pi XRD peaks are observed in the Co–Pi/$BiVO_4$ film.

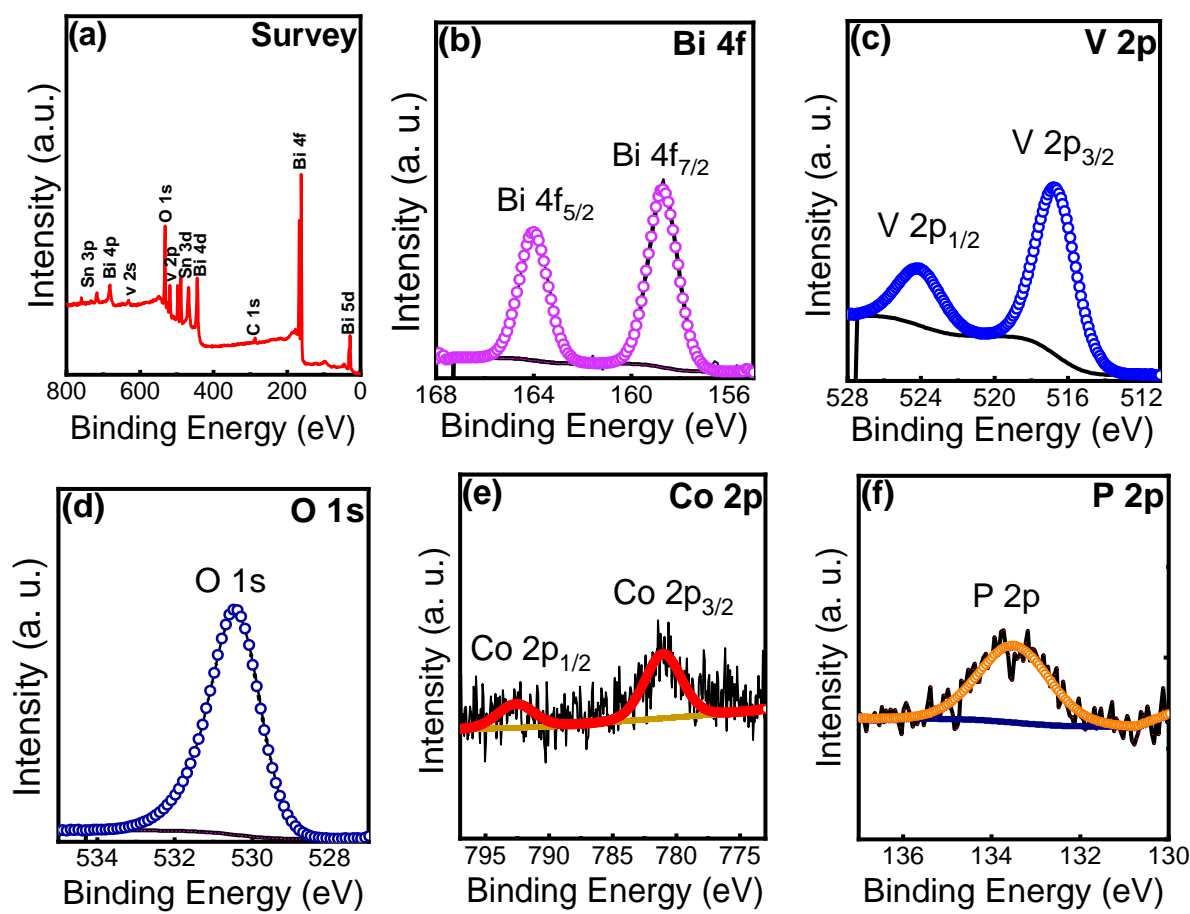

**Figure S2.** (a) XPS survey spectrum of the 270-nm thick BiVO$_4$ film after etching by Ar$^+$ ions sputtering for 120 seconds. The spectrum reveals photoemission peaks corresponding to Bi, V, and O, as well as an Sn peak arising from the FTO substrate. The presence of the strong V peaks in the XPS spectrum confirms the intercalation of V ions throughout the thickness of the Bi$_2$O$_3$ film, extending down to the substrate. (b) XPS Bi 4f peaks showing a binding energy difference of 5.3 eV between Bi 4f$_{7/2}$ and Bi 4f$_{5/2}$ spin-orbit components, providing evidence for the oxidation state of Bi$^{3+}$.[1] (c) XPS V 2p spectrum revealing the V$^{5+}$ oxidation state with distinct spin-orbit peaks at 516.7 eV (V 2p$_{3/2}$) and 524.2 eV (V 2p$_{1/2}$).[1] (d) XPS O 1s spectrum at 530.4 eV associated with lattice oxygen. For the Co–Pi/BiVO$_4$ film, the Co–Pi co-catalyst integration on the film surface is confirmed by the presence of (e) Co 2p peaks at 781.1 eV (Co 2p$_{3/2}$) and 793.2 eV (Co 2p$_{1/2}$), along with (f) the P 2p peak at 133.4 eV.[2]

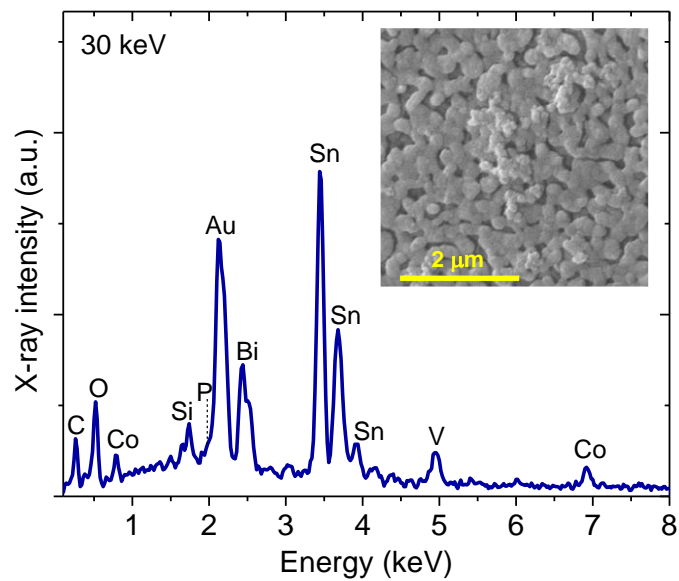

**Figure S3.** EDX spectrum of the Co–Pi/BiVO$_4$ film revealing the presence of all the anticipated elements: Bi, V, O, Co and P. The peaks of Sn, Si, and Au arise the FTO glass substrate and the gold coating used for SEM imaging purposes. The SEM image (inset) shows Co-Pi nanoparticles on the BiVO$_4$ film.

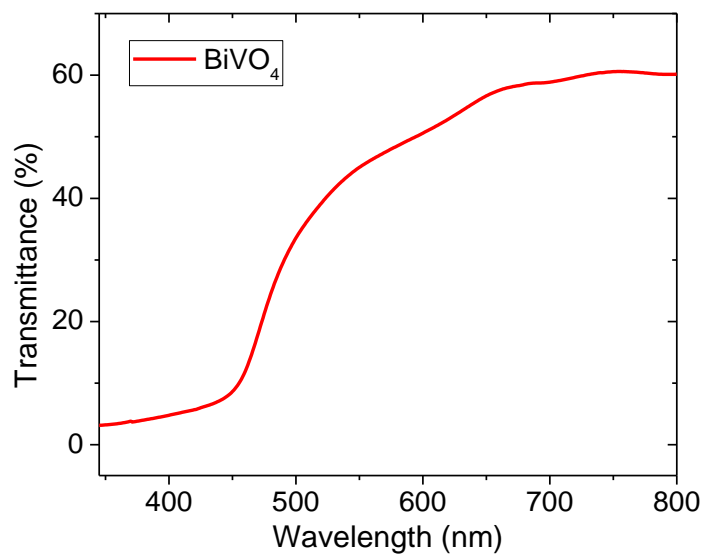

**Figure S4**. Optical transmission spectrum of the optimized BiVO$_4$ film showing a transmittance of < 10% for wavelengths below 450 nm and a broad edge that allows this film to absorb well into the visible range.

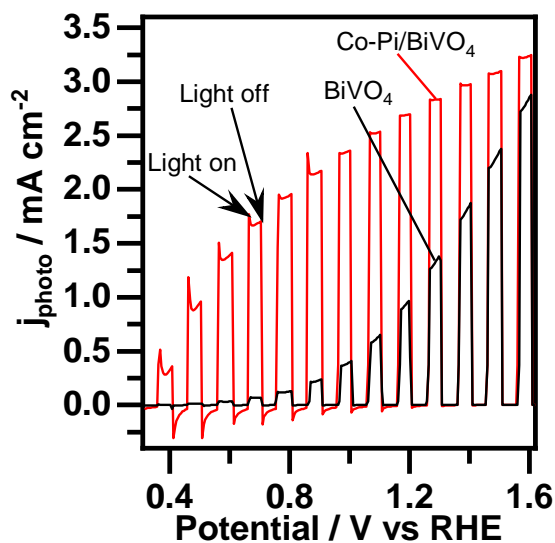

**Figure S5.** Chopped LSV curves showing the photoelectrochemical water oxidation of the BiVO$_4$ and Co–Pi/BiVO$_4$ photoanodes, measured in 0.5 M PBS solution under AM 1.5G illumination. The photocurrent displays a notable enhancement after the integration of Co–Pi co-catalyst, rising from 1.0 mA/cm$^2$ in the BiVO$_4$ photoanode to 2.9 mA/cm$^2$ in the Co–Pi/BiVO$_4$ configuration at 1.23 V$_{RHE}$. The LSV curves exhibit a diminishing trend in anodic spikes with increasing applied potential, attributed to enhanced charge separation at higher potentials.

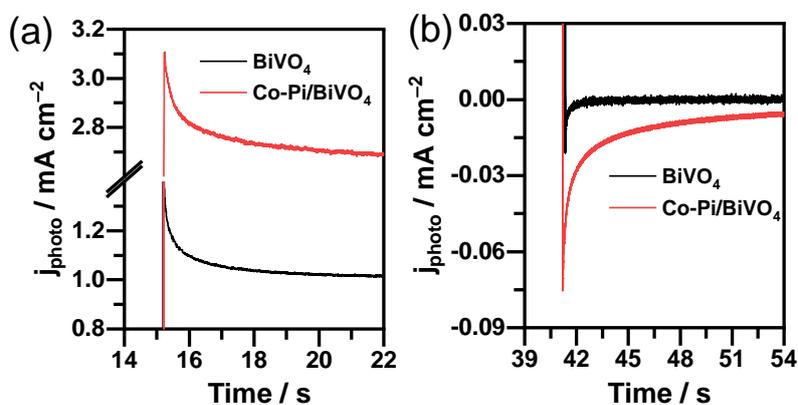

**Figure S6.** (a) Anodic and (b) Cathodic spikes observed for the BiVO$_4$ and Co–Pi/BiVO$_4$ photoanodes during transient photocurrent measurements at 1.23 V$_{RHE}$ in 0.5 M PBS solution under AM 1.5G illumination. The photocurrent transient of the Co–Pi/BiVO$_4$ configuration exhibits a notably heightened cathodic spike when switching off the light, indicating increased hole accumulation at the surface. This result is consistent with the suppression of surface recombination due to the Co–Pi catalyst.

**Table S1:** Comparison of PEC performance at 1.23 $V_{RHE}$ for BiVO$_4$ photoanodes in this work and recent reports. To ensure a fair comparison, the table is limited to undoped BiVO$_4$ photoanodes with and without Co-Pi co-catalysts, measured in a neutral electrolyte and under conditions similar to those employed in this work.

| Synthesis method | Electrolyte (pH = 7) | J (mA/cm$^2$) at 1.23 $V_{RHE}$ | | Reference |
|---|---|---|---|---|
| | | No catalyst | Co-Pi co-catalyst | |
| V intercalation of Bi$_2$O$_3$ | 0.5 M H$_2$KO$_4$P | 1.0 | 2.9 | This work |
| Electrodeposition | 0.1 M KPi | 0.5 | 0.8 | [3] |
| Drop-casting | 0.5 M H$_2$KO$_4$P | 0.4 | 2.3 | [4] |
| Calcination of Bi$_2$O$_3$ | 0.5 M H$_2$KO$_4$P | 2.6 | 2.9 | [5] |